\documentclass[aps,pra,twocolumn,superscriptaddress]{revtex4-1}
\usepackage{graphicx}
\usepackage{amssymb}
\usepackage{amsmath}
\usepackage{bm}
\usepackage{xcolor}

\begin{document}

\title{Metastable hard-axis polar state of a spinor Bose-Einstein condensate \\under a magnetic field gradient}
\author{Joon Hyun Kim}
\affiliation{Department of Physics and Astronomy, and Institute of Applied Physics, Seoul National University, Seoul 08826, Korea}

\author{Deokhwa Hong}
\affiliation{Department of Physics and Astronomy, and Institute of Applied Physics, Seoul National University, Seoul 08826, Korea}
\affiliation{Center for Correlated Electron Systems, Institute for Basic Science, Seoul 08826, Korea}

\author{Seji Kang}
\affiliation{Department of Physics and Astronomy, and Institute of Applied Physics, Seoul National University, Seoul 08826, Korea}
\affiliation{Center for Correlated Electron Systems, Institute for Basic Science, Seoul 08826, Korea}

\author{Y. Shin}
\email{yishin@snu.ac.kr}
\affiliation{Department of Physics and Astronomy, and Institute of Applied Physics, Seoul National University, Seoul 08826, Korea}
\affiliation{Center for Correlated Electron Systems, Institute for Basic Science, Seoul 08826, Korea}

\date{\today}

\begin{abstract}

We investigate the stability of a hard-axis polar state in a spin-1 antiferromagnetic Bose-Einstein condensate under a magnetic field gradient, where the easy-plane spin anisotropy is controlled by a negative quadratic Zeeman energy $q<0$. In a uniform magnetic field, the axial polar state is dynamically unstable and relaxes into the planar polar ground state. However, under a field gradient $B'$, the excited spin state becomes metastable down to a certain threshold $q_{th}$ and as $q$ decreases below $q_{th}$, its intrinsic dynamical instability is rapidly recalled. The incipient spin excitations in the relaxation dynamics appear with stripe structures, indicating the rotational symmetry breaking by the field gradient. We measure the dependences of $q_{th}$ on $B'$ and the sample size, and we find that $q_{th}$ is highly sensitive to the field gradient in the vicinity of $B'=0$, exhibiting power-law behavior of $|q_{th}|\propto B'^{\alpha}$ with $\alpha \sim 0.5$. Our results demonstrate the significance of the field gradient effect in the quantum critical dynamics of spinor condensates.

\end{abstract}

\maketitle

\section{Introduction}

Bose-Einstein condensates (BECs) of atoms with internal spin degree of freedom, so-called spinor BECs, represent a quantum fluid system with both superfluidity and magnetic order~\cite{Ho98,Machida98,Ueda12,Stamper-Kurn13}. Owing to their rich phase diagram and the experimental capability of dynamically controlling the system parameters, the spinor BECs provide unique opportunities for studying quantum phase transition dynamics, which is an important subject to understand in modern physics~\cite{Vengalattore11,Eisert15}. For the study, a quantum quench protocol is typically employed: a system is initially prepared in a well-defined phase and, by changing the system's Hamiltonian, it is driven to undergo a phase transition into another phase with different symmetry. As the evolution of the system's wave function is directly probed, many aspects of the phase transition dynamics can be revealed in an unprecedented level of details, including domain and defect formation~\cite{Stamper-Kurn06,Stamper-Kurn11,Spielman14}, phase-ordering dynamics~\cite{DasSarma14,Blakie16,Blakie17,Blakie18}, and possible dynamic scaling behavior~\cite{Oberthaler15,Chapman16,Davis11,Matuszewski13}.

Recently, a quantum phase transition from the easy-axis polar (EAP) phase to the easy-plane polar (EPP) phase was studied with spin-1 antiferromagnetic BECs~\cite{Raman11,Raman17,Kang17}. For antiferromagnetic interactions, the ground spin state is a polar state with $\langle\textbf{F}\rangle=0$, where $\textbf{F}$ is the spin-1 operator for the atoms, and the order parameter is represented by a unit vector $\boldsymbol{d}$ for spin orientation such that the BEC is in the $m_F=0$ state for the quantization axis parallel to $\boldsymbol{d}$. In the presence of an external magnetic field, e.g., along the $z$ direction, the system has uniaxial spin anisotropy due to the quadratic Zeeman energy $E_q= q \langle F_z^2\rangle$, where $F_z$ is the $z$ projection of $\textbf{F}$. Depending on the sign of $q$, there are two phases for the ground state: for $q>0$, EAP phase with $\boldsymbol{d} \parallel\hat{z}$ and for $q<0$, EPP phase with $\boldsymbol{d} \perp \hat{z}$. The EAP-to-EPP phase transition is special in that, although it is first-order, metastable states are prohibited due to the flat energy landscape at the quantum critical point $q_c=0$~\cite{Ueda13}. The quench dynamics from the EAP phase to the EPP phase was experimentally investigated by suddenly changing the $q$ value from positive to negative~\cite{Raman11,Raman17, Kang17}. It was shown that the initial axial polar state with $\boldsymbol{d}=\hat{z}$ is dynamically unstable for $q<0$ and spin fluctuations are exponentially amplified to cause the phase transition. In a 2D situation~\cite{Kang17}, the quench dynamics was characterized by the emergence and decay of spin turbulence and the formation of topological defects, such as half-quantum vortices~\cite{Seo15}.

A notable observation in the previous experiments was that the dynamical instability of the initial axial polar state is significantly suppressed by a magnetic field gradient~\cite{Raman17}. In the early works of antiferromagnetic spinor BECs~\cite{Ketterle98,Ketterle99}, the metastability of the axial polar state under a field gradient was demonstrated for $q>0$, where the ground state of the system has an inhomogeneous spatial structure of spin domains. So, it is natural to ask how the suppression of dynamical instability on the $q<0$ side is related to the metastability on the $q>0$ side. In prospect of the quantitative study of scaling behavior near the critical point, it is particularly important to understand the field gradient effect on the phase transition dynamics.  

In this paper, we examine the stability of the axial polar state in the antiferromagnetic BEC under an external magnetic field gradient. In contrast to the previous result~\cite{Raman17}, we observe that the dynamical instability is {\it fully} suppressed by a magnetic field gradient down to a certain threshold value of $q_{th}<0$, demonstrating the metastability of the axial polar state in the $q<0$ region with easy-plane spin anisotropy. Furthermore, we observe that below the threshold $q_{th}$, the dynamical instability is rapidly recalled in the system. The incipient spin excitations in the relaxation dynamics exhibit a stripe spatial structure, indicating the rotational symmetry breaking by the field gradient. The transition from the metastable regime to the dynamically unstable regime occurs quite suddenly, and the threshold value $q_{th}$ is unambiguously determined. We measure $q_{th}$ as a function of the field gradient $B'$ as well as the sample size, and find that $q_{th}$ is highly sensitive to $B'$ at $B'=0$. Our results show that the field gradient effect is significant in the quantum quench dynamics of the spinor BEC. This is critical in the efforts to precisely determine the properties near the quantum critical point and to probe possible scaling behavior in the quantum phase transition dynamics.

This paper is organized as follows. In Sec.~II, we present a brief description of the mean-field ground state under a magnetic field gradient. In Sec.~III, we describe our quantum quench experiment, and in Sec.~IV, we present the experimental results. Finally, summary and outlook are provided in Sec.~V.

\section{Mean-field ground state}

In a mean-field description, the spin-dependent energy functional of a spin-1 spinor BEC is given by
\begin{equation}
E_{spin}=\int\mathrm{d}{\bf r}~n\Big(\frac{c_2 n}{2}\langle\textbf{F}^2\rangle-(p-p_0)\langle F_z\rangle+q\langle F_z^2\rangle\Big)
\end{equation}
where $n({\bf r})$ is the density of atoms~\cite{Ueda12,Stamper-Kurn13}. The first term is the spin interaction energy and $c_2>0$ for antiferromagnetic interactions. The second and third terms are, respectively, the linear and quadratic Zeeman energies for an external magnetic field $B(\bf r)$ along the $z$ direction. Here, $p({\bf r})=g_F\mu_B B({\bf r})$, with $g_F$ being the Land\'{e} hyperfine $g$-factor and $\mu_B$ the Bohr magneton, and $p_0$ is the Lagrange multiplier for the  total magnetization $M_z=\int \mathrm{d}{\bf r}~n \langle F_z \rangle$, which is conserved in the system. For $M_z=0$, the magnetic field gradient effect is represented by $\tilde{p}(\textbf{r}) \equiv p-p_0=g_F\mu_B B'x$, where the field gradient direction is assumed to be the $x$ direction.  

The order parameter of the BEC is expressed as $\Psi=(\psi_1,\psi_0, \psi_{-1})^{\mathrm{T}}$, where $\psi_{l}$ is the $m_z=l$ spin component of the condensate ($l=1,0,-1$), and its ground state is given by
\begin{equation}
\Psi_\textrm{I}=\sqrt{n} e^{i\theta}
\begin{pmatrix} 0\\1\\0 \end{pmatrix}
\end{equation}
for $\tilde{p}^2<2 c_2 n q$ and
\begin{equation}
\Psi_\textrm{II}=\sqrt{\frac{n}{2}} e^{i\theta}
\begin{pmatrix} -e^{-i\phi}\sqrt{1+f_z} \\ 0\\ e^{i\phi}\sqrt{1-f_z} \end{pmatrix}
\end{equation}
 for $\tilde{p}^2> 2 c_2 n q$, where $\theta$ is the superfluid phase and  $f_z=\min[\frac{\tilde{p} }{c_2 n},1]$~\cite{Ueda12}. $\Psi_\textrm{I}$ is the axial polar state having only the $m_z=0$ component and $\Psi_\textrm{II}$ is a mixture of the $m_z=\pm1$ components with net magnetization $f_z$. For $f_z=0$, $\Psi_\textrm{II}$ corresponds to the planar polar state with $\boldsymbol{d}=\cos \phi~\hat{x} + \sin \phi~\hat{y}$. 

The spatial structure of a BEC trapped in a harmonic potential and under a magnetic field gradient $B'$ can be described in the local density approximation. As $\tilde{p}$ and $n$ vary spatially over the sample, different spin domains are formed in the trapped condensate and their interface locates at $\textbf{r}_c$ with $x_c=\pm\frac{\sqrt{2c_2 n (\textbf{r}_c) q}}{g_F \mu_B B'}$, which is determined from $\tilde{p}^2=2c_2 nq$. For $q>0$, an axial polar domain of the $m_z=0$ component is formed in the inner region of $|x|<|x_c|$ and magnetized domains, having an unequal mixture of the $m_z=\pm 1$ components, appear in the outer region of $|x|>|x_c|$ [Fig.~1(a)II].

\begin{figure}[t]
	\includegraphics[width=8.5cm]{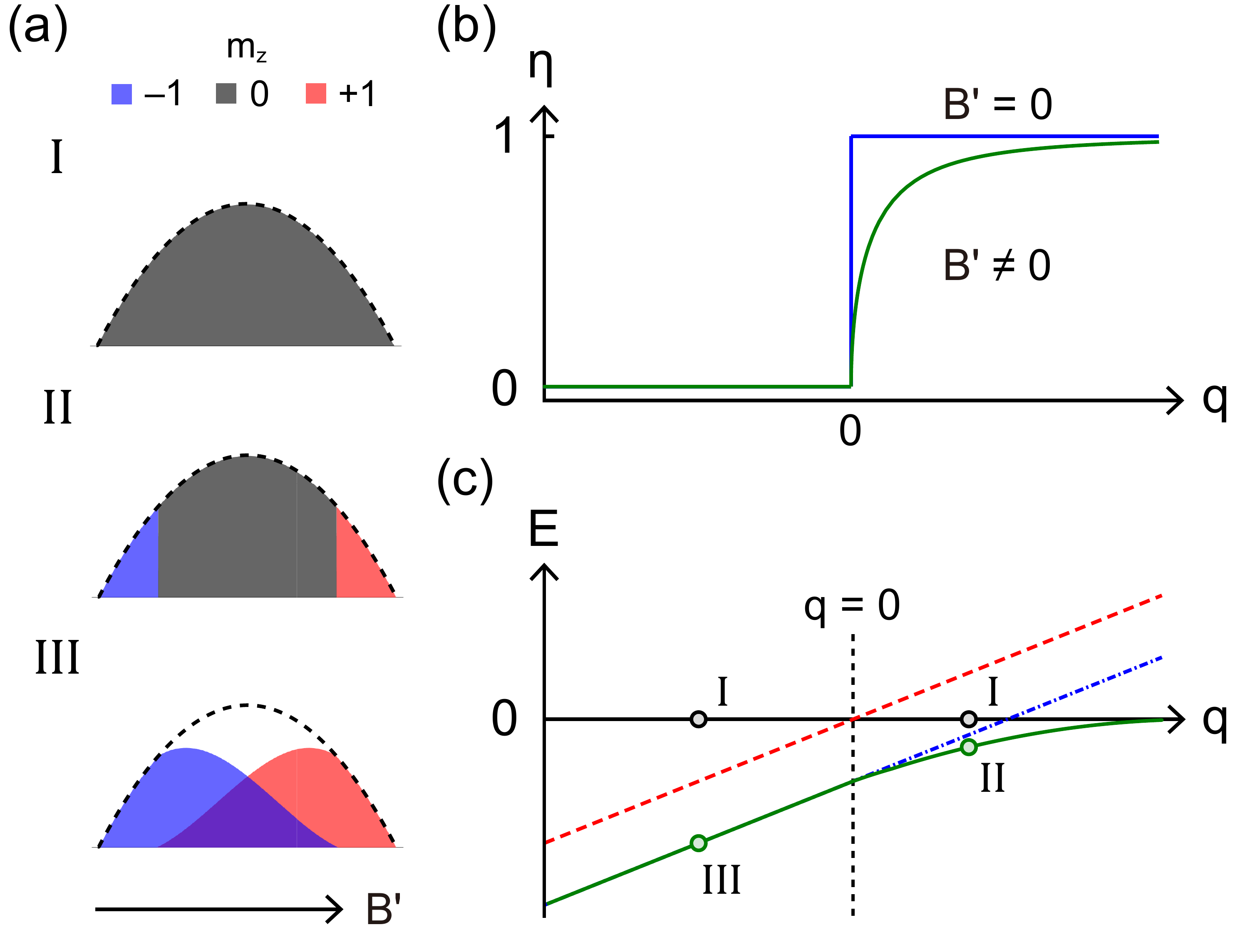}
	\caption{Spin-1 antiferromagnetic Bose-Einstein condensate (BEC) in a harmonic trapping potential under a magnetic field gradient $B'$. (a) Schematic description of the spatial structure of the BEC in various states: (I) axial polar state and (II) ground states for positive quadratic Zeeman energy $q>0$ and (III) for $q<0$. The blue, grey, and red shades represent the $m_z=-1,0,1$ spin components, respectively, and the black dashed line indicates the density of the condensate. (b) Fractional population $\eta$ of the $m_z=0$ component in the ground state of the trapped condensate as a function of $q$ for $B'=0$ (blue line) and $B'\neq0$ (green line). (c) Energy level of the BEC in the ground state as a function of $q$ (green line). The $E=0$ horizontal line denotes the energy level of the axial polar state, and the red dashed line shows the energy level of the planar polar state. In this work, we investigate the relaxation dynamics from the axial polar state (I) to the ground state (III) for $q<0$ in the presence of the magnetic field gradient.}
\end{figure}

In Fig.~1(b), the fractional population $\eta$ of the $m_z=0$ component in the trapped BEC is displayed as a function of $q$. Here, we assume a Thomas-Fermi (TF) density profile of $n(\mathbf{r})= \max[n_0\big(1-\frac{x^2}{R_x^2}-\frac{y^2}{R_y^2}-\frac{z^2}{R_z^2}\big),0]$ and calculate $\eta$ by integrating the density over the inner region bounded by the domain interface, where we obtain the analytic expression $\eta_m(q,B')=\frac{\beta}{2}\frac{2\beta^2 + 3R_x^2}{(\beta^2 + R_x^2)^{3/2}}$ with $\beta=\frac{\sqrt{2c_2 n_0 q}}{g_F\mu_B B'}$ for $q>0$. As $q$ decreases below zero, $\eta$ vanishes, i.e., the axial polar domain in the middle of the trapped condensate shrinks to disappear, and for $q<0$, the ground state becomes a two-component BEC, where the $m_z=\pm 1$ components are displaced symmetrically along the direction of a field gradient [Fig.~1(a)III]. In the case of zero field gradient, $\eta$ shows a step change from 1 to 0 at $q=0$, which represents the quantum phase transition between the EAP and EPP phases.

In Fig.~1(c), we plot the energy level of the ground state under a magnetic field gradient as functions of $q$, together with those of the axial polar state and the planar polar state. To introduce the current work in comparison with the previous experiments, we mark several points of interest in the plot by I, II, and III. In Ref.~\cite{Ketterle99}, it was demonstrated that the axial polar state (I) in the $q>0$ region is metastable against relaxation to the ground state with spin domains (II). In Refs~\cite{Kang17,Raman17}, it was observed that the axial polar state (I) in the $q<0$ region is dynamically unstable to relax into the planar polar state (dashed red line) for $B'\approx 0$. In this work, we examine the stability of the axial polar state in the $q<0$ region for relaxation into the ground state under a magnetic field gradient (III).

\section{Experiment}

We start our experiment by preparing a BEC of $^{23}$Na atoms in the $|F=1,m_{F}=0\rangle$ hyperfine spin state in a highly oblate optical dipole trap (ODT) with the trapping frequency of $(\omega_{x},\omega_{y},\omega_{z}) = 2\pi\times(5.1,7.2,548)$~Hz. The sample preparation is carried out in a magnetic field of $B_z=0.2$~G. The condensate contains typically $N_{tot}\approx 4.5\times10^{6}$ atoms and its TF radii are given by $(R_{x},R_{y},R_{z})\approx(186,131,1.7)~\mu$m. The density and spin healing lengths are given by, respectively, $\xi_{n}=\hbar/\sqrt{2mc_{0} n_0}\approx 0.5~\mu$m and $\xi_{s}=\hbar/\sqrt{2mc_2 n_0}\approx 3.8~\mu$m for the peak atomic density $n_0$ of the condensate, where $m$ is the atomic mass and $c_{0 }$($c_2$) denotes the density (spin) interaction coefficient~\cite{Lett07}. Note that $R_{z}< \xi_{s} \ll R_{x,y}$ and spin dynamics in our system are effectively 2D. In the experiment, $N_{tot}$ is controlled by adjusting the evaporation cooling process in the ODT and the final thermal fraction of the sample is kept less than $15~\%$.

A quantum quench experiment is performed by suddenly changing the sign of $q$ with a microwave field dressing technique, as described in Ref.~\cite{Kang17}. First, we adiabatically ramp $B_z$ down to 50~mG. During the field ramp, we control a magnetic field gradient $B'=\frac{\partial |B_z|}{\partial x}$ along the $x$ direction. $B'$ is calibrated using an interferometric technique, where the growth rate of the spatial frequency of Ramsey interference fringes between the spin components is measured~\cite{Hirano13}. At $B_z=50$~mG, $q/h\approx 0.7$~Hz and the initial $m_z=0$ state is dynamically stable under the field gradient~\cite{Ketterle99}.  Then, we suddenly turn on a microwave field to tune $q$ to a target value $q_f$~\cite{Bloch06,Liu14}. After a variable hold time $t$, we take an absorption image of the sample with Stern-Gerlach (SG) spin separation. After releasing the trapping potential and turning off the microwave field, a short pulse of an additional magnetic field gradient is applied to spatially separate the three $m_z=-1,~0,~1$ spin components and the magnetic field is ramped to $B_z=0.5$~G during a 24-ms time of flight for imaging.

\section{Results}

\subsection{Metastable axial polar state}

\begin{figure}[t]
	\includegraphics[width=8.5cm]{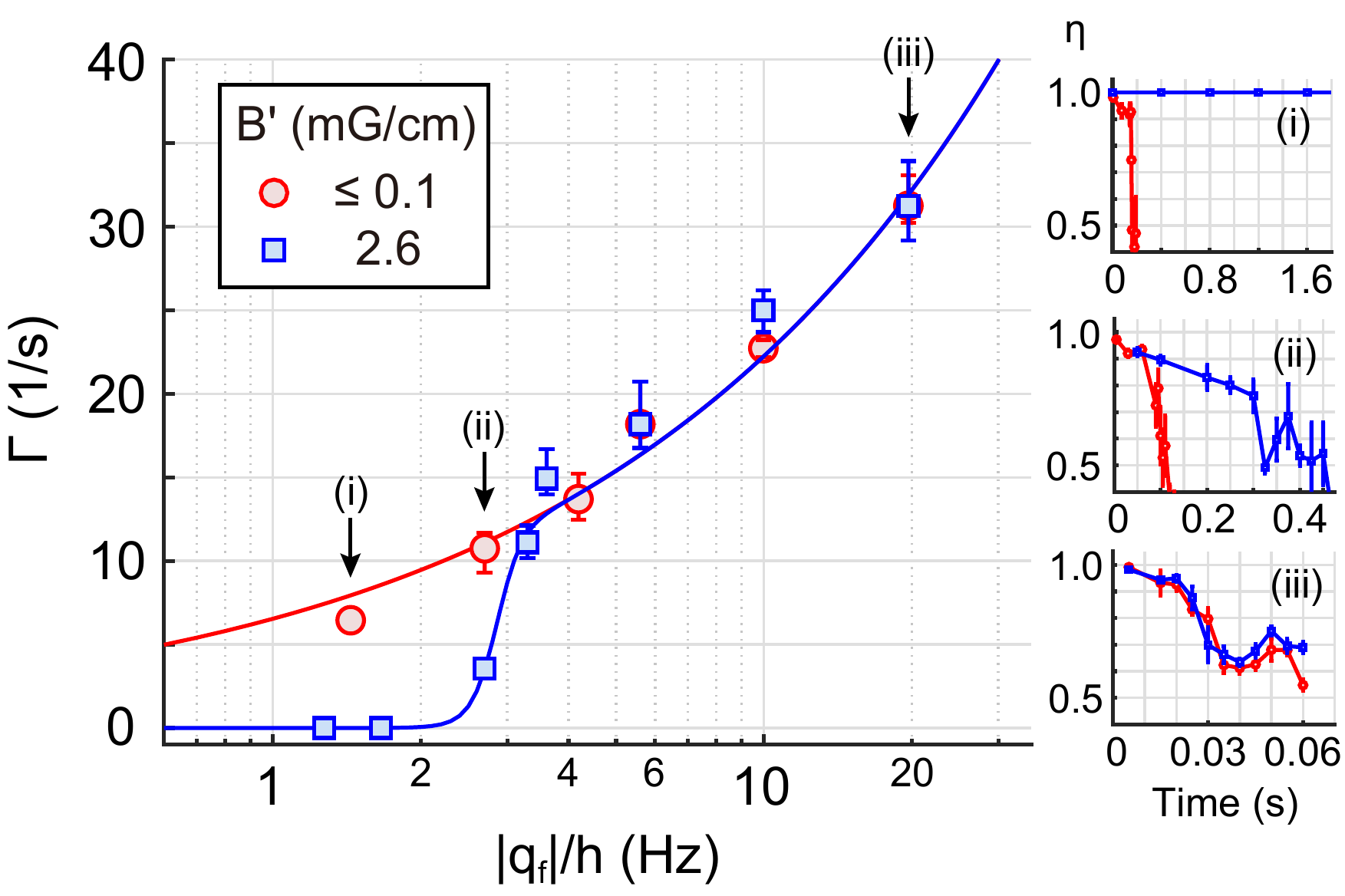}
	\caption{Decay rate $\Gamma$ of the axial polar state after the quench to $q_f<0$ as a function of $|q_f|/h$. The red circles and the blue squares show $\Gamma$ for the $B'\approx 0$ and 2.6~mG/cm, respectively. $\Gamma$ is obtained from the curve fit of $\eta(t)=1-\alpha e^{\gamma t}$ to the experimental data (right insets) and its error bar indicates a 95\% confidence interval of the fit. The red solid line denotes a power-law fit to the $B'\approx0$ data, which gives an exponent of 0.53(10). The blue solid line is a guide line for eyes to the $B'\approx 2.6$~mG/cm data, which shows  onset behavior for $|q_f|/h>2$~Hz. The right insets display the temporal evolutions of $\eta$ at (i) $q_f/h\approx-1.4~$Hz, (ii) $-2.7~$Hz, and (iii) $-20~$Hz for $B'\approx 0$ (red) and 2.6~mG/cm (blue). Each data point in the insets was typically obtained from five measurements of the same exeperiments and the standard error of the mean is denoted by its error bar.}
\end{figure}

We measure the temporal evolution of $\eta$ to characterize the stability of the axial polar state after the quench to $q_f<0$. The decay rate $\Gamma$ is determined as the inverse of the time $t_0$ at which $\eta(t_0)=0.7$, where $t_0$ is obtained from a curve fit of $\eta(t)=1-\alpha e^{\gamma t}$ to the experimental data for $\eta>0.6$ at the initial stage. Figure 2 displays the measurement results of $\Gamma$ for $B'\approx 0$ and $2.6$~mG/cm as a function of $|q_f|/h$. For $B'\approx 0$, $\Gamma$ exhibits a power-law behavior of $\Gamma \propto |q_f|^{\gamma}$ with an exponent of $0.53(10)$, which is consistent with the result of the previous experiment~\cite{Kang17}. For $B'\approx 2.6$~mG/cm, $\Gamma$ is found to be suppressed in a small $|q_f|$ region and as $|q_f|$ increases, $\Gamma$ is restored to the value for $B'\approx0$. Remarkably, for $|q_f|/h<2$~Hz, it was observed that $\eta$ remains unity for a long hold time, even up to 10~s, which is limited by the sample lifetime under the microwave field dressing. In this case, $\Gamma$ cannot be determined and we practically set $\Gamma=0$. This observation clearly shows that the dynamical instability of the axial polar state is not only mitigated by a magnetic field gradient, but completely quenched to make the spin excited state metastable against relaxation into the ground state.

Since in Ref.~\cite{Ketterle99} the axial polar state was shown to be metastable under a field gradient for $q>0$, our observation seems to be a simple extention of its metastability to the $q<0$ region. However, we note that the underlying mechanism for negative $q$ should be different from that for positive $q$, because of the absence of the energy barrier for local spin relaxation. For $q>0$, local spin relaxation via spin-exchange process is prohibited due to the energy barrier of the positive quadratic Zeeman energy and requires thermal activation. However, for $q<0$, it is an exothermic process that would necessarily enhance the relaxation of the axial polar state. Furthermore, one can think of an evolution path of the system from the initial excited state to the ground state along which the system's energy continuously decreases. For example, the spin is first rotated from $\boldsymbol{d}=\hat{z}$ to $\boldsymbol{d}=\hat{y}$, which reduces the quadratic Zeeman energy, and the resulting planar polar state is transformed to the ground state by redistributing the $m_z=\pm 1$ spin components in the trapped condensate. Therefore, the metastability observed in the $q<0$ region cannot be accounted for by the conventional frame based on energy barrier in the system configuration.

\begin{figure}[t]
	\includegraphics[width=8.5cm]{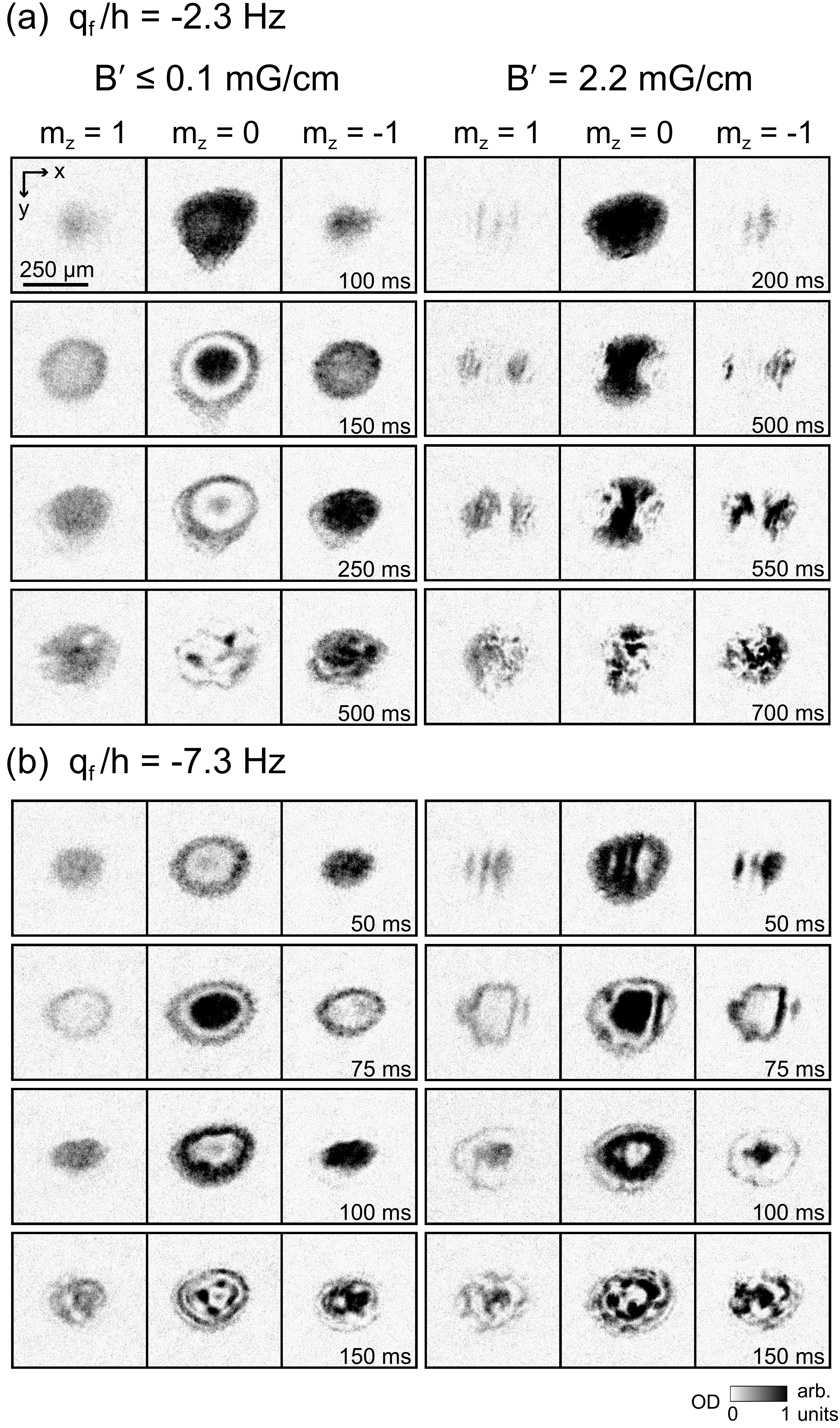}
	\caption{Field gradient effect on the quench dynamics of a BEC. Absorption images of the three $m_z=-1,0,1$ spin components for various hold times after the quench to (a) $q_f/h\approx -2.3$~Hz and (b) $-7.3$~Hz. For $B'\approx 2.2$~mG/cm, $q_f/h\approx -2.3$~Hz is in the transition between the metastable and dynamically unstable regimes and the condensate relaxes slowly in comparison to that for $B'\approx 0$. In the images, the optical densities (ODs) of the spin components appear different due to their different optical transition strengths in the $F=1$ absorption imaging~\cite{Kim16}. }
\end{figure}

We indeed observe that the metastable axial polar state is fragile for small spin perturbations. When the spin director $\boldsymbol{d}$ is tilted slightly away from the $z$ axis by applying a short pulse of radio-frequency (RF) field, the BEC immediately proceeds to relax into the ground spin state. Also, when the system has small ferromagnetic domains at its both ends, which is achieved by preparing the sample in the ground state with positive $q$ under a small magnetic field gradient, the spin domains at both ends permeate through the condensate after the quench to $q_f<0$, resulting in the ground spin state. This indicates that the metastable hard-axis polar state is sustained as a global property of the condensate. Meanwhile, we find that the metastable spin state is intact in a turbulent condensate containing many vortices, which is prepared by violently stirring the sample with an optical obstacle~\cite{Kwon14,Seo17}, demonstrating its stability for density perturbations.

\subsection{Recurrence of dynamical instability}

As $|q_f|$ increases, the metastability of the axial polar state eventually breaks down (Fig.~2). The rapid recovery of $\Gamma$ to the value of $B'\approx 0$ and the resemblance of the evolution curve $\eta(t)$ for high $|q_f|$ suggest that the system suffers from the same dynamical instability as in the $B'\approx 0$ case, which is associated with transverse magnon excitations. However, the details of the relaxation dynamics under a field gradient are observed to be different from those for $B'\approx 0$. In Fig.~3, we display absorption images of the sample at the early stage of the quench evolution for $B'\approx 0$ and 2.2~mG/cm. At $B'\approx 0$, the quench dynamics proceeds first with generating ring-shaped, long-wavelength spin excitations, whose spatial pattern is inherited from the elliptical geometry of the condensate~\cite{Kang17}. On the other hand, when a field gradient is applied, the incipient spin excitations appear with a stripe structure that is orthogonal to the field gradient direction and has higher wavenumber. This means that the most unstable modes for the dynamical instability are modified in the presence of the field gradient, which breaks the rotational symmetry of the system. At sufficiently high $|q_f|$, the stripe spin excitations are followed by ring-shaped excitations [Fig.~3(b)], implying tight competition between the two excitation modes. In the subsequent relaxation, an irregular spin texture develops, wherein global spin polarization gradually develop over the condensate, according to the applied field gradient.

In Fig.~4, we display the fractional population of the $m_z=0$ component, $\eta$, measured at $t=3$~s as a function of $q_f/h$ for various $B'$. $\eta$ collapses rapidly as $q_f$ decreases below a threshold, which represents the sharp transition from the metastable regime to the dynamically unstable regime, as observed in Fig.~2. The sudden collapse behavior allows unambiguous determination of the threshold value $q_{th}$, which we determine using a sigmoid function fit to the experimental data. With increasing $B'$, the metastable-to-unstable transition position is further shifted away from the critical point $q_c=0$. This corroborates that the field gradient induces the metastability of the axial polar state.

\subsection{Dependence of $q_{th}$ on $B'$ and $N_{tot}$}

\begin{figure}[t]
	\includegraphics[width=8.5cm]{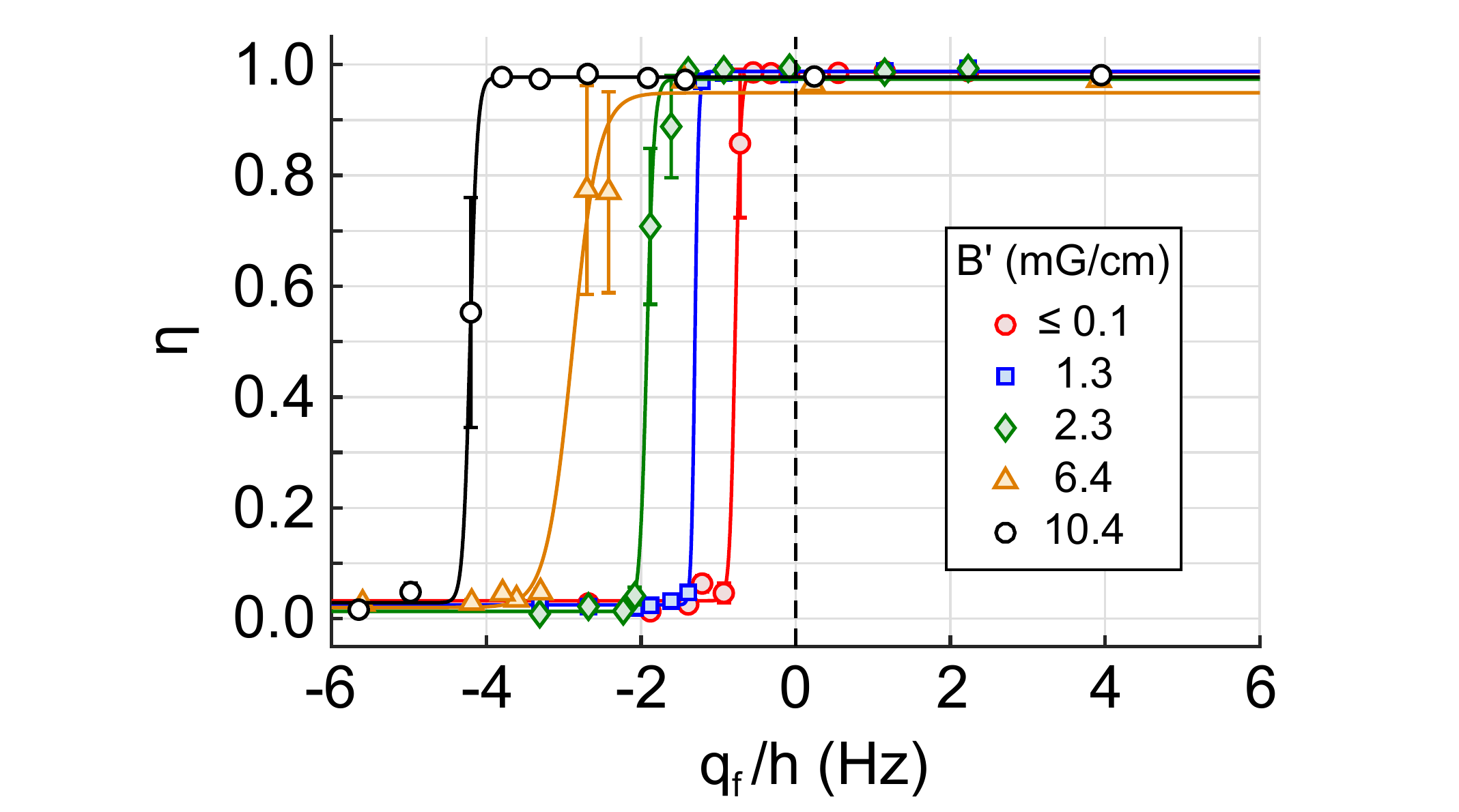}
	\caption{Fractional population $\eta$ of the $m_z=0$ component measured at a hold time $t=3$~s after the quench as a function of $q_f/h$, for various $B'$ ($N_{tot}\approx 4.5\times 10^6$). The solid lines denote sigmoid fits to each measurement data set, from which the threshold $q_{th}$ is determined for the collapse of the metastable hard-axis polar state. Each data point was typically obtained from five measurements and its error bar indicates the standard error of the mean.}
\end{figure}

\begin{figure}[t]
	\includegraphics[width=8.5cm]{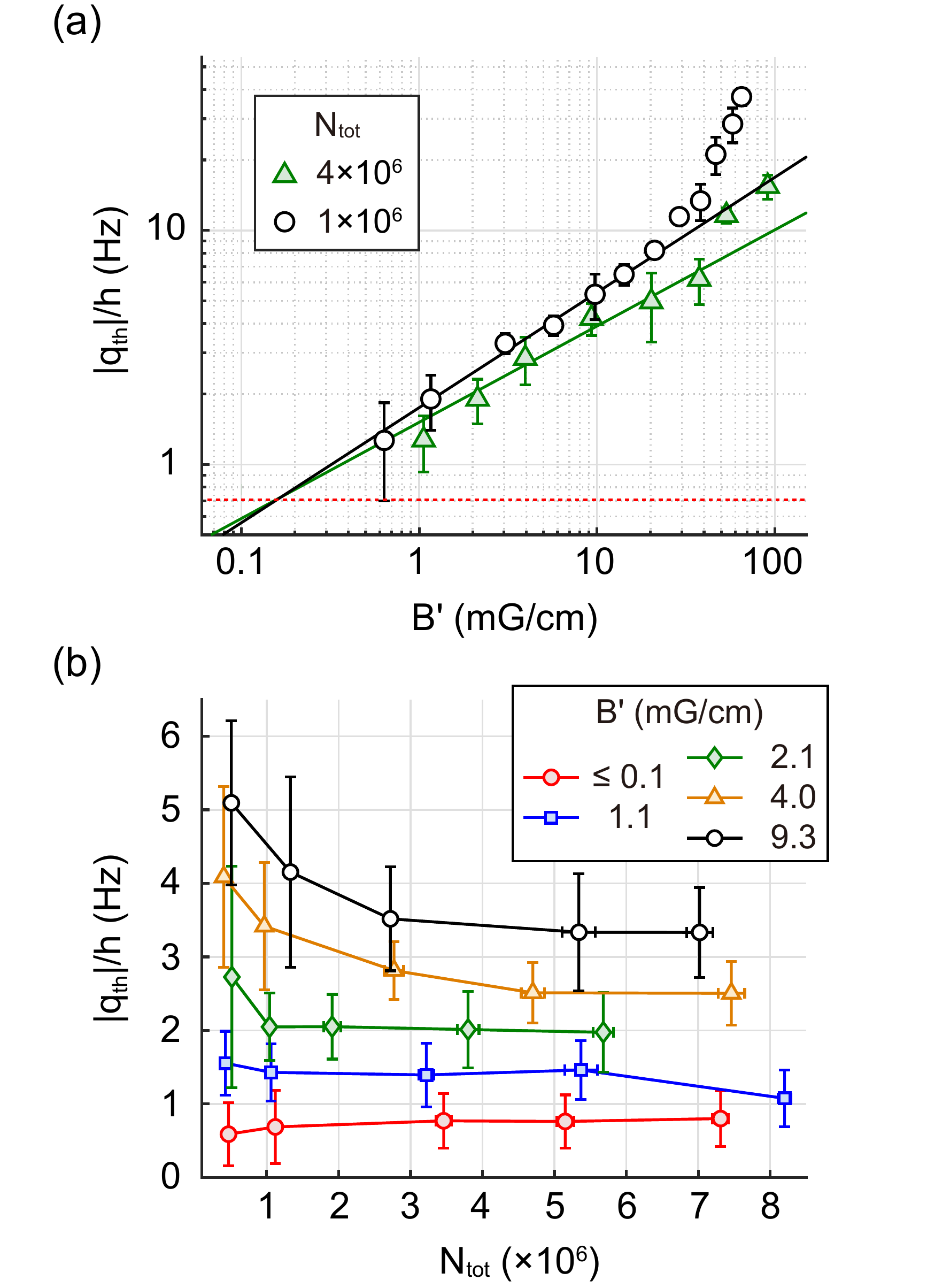}
	\caption{Dependence of $q_{th}$ on $B'$ and $N_{tot}$. (a) $|q_{th}|/h$ as a function of $B'$. The green triangles (black open circles) display the measurement results with $N_{tot}\approx 4.0\times10^6$ ($1.0\times10^6$).  The solid lines denote power-law fits to each data set up to $B'\approx20$~mG/cm, where the exponents are given by $\alpha=0.41(18)$ (green) and $\alpha=0.49(07)$ (black), respectively. The red dotted line indicates the $|q_{th}|/h$ value measured for $B'\leq0.1$~mG/cm. (b) $|q_{th}|/h$ as a function of $N_{tot}$ for varous $B'$. The error bar of each data point indicates the range of $0.1<\eta<0.9$ in the sigmoid curve fit to the experimental data of $\eta$ (Fig.~4), including the long-term systematic drift of $\pm0.3~$Hz in our measurements.}
\end{figure}

We investigate the dependence of $q_{th}$ on $B'$ over a broad range of field gradients up to $B'<$~90~mG/cm [Fig.~5(a)]. $|q_{th}|$ monotonically increases with growing $B'$. Remarkably, for low $B'$, the increase of $|q_{th}|$ exhibits power-law behavior. We characterize the behavior by fitting a curve of $|q_{th}|/h=\kappa B'^\alpha$ to the measurement results for $B'\lesssim 20$~mG/cm, obtaining $\alpha=0.41(18)$ and 0.49(07) for $N_{tot}\approx4\times10^6$ and $1\times10^6$, respectively. As $B'$ increases over a few tens of mG/cm, $q_{th}$ deviates from the power-law relation with $B'$. Interestingly, we notice that the magnitudes of the field gradient at which the deviation becomes noticeable, are close to the values given by $\mu_B B' R_x=\mu$, which are $B'\approx 38$~mG/cm and $B'\approx 29$~mG/cm for, respectively, $N_{tot}\approx 4\times10^6$ and $1\times10^6$, where $\mu=\frac{1}{2}m\omega_x^2 R_x^2$ is the chemical potential of the condensate.

The field gradient effect of suppressing the dynamical instability might be described as a result of spatial dephasing of the spin excitations, which are otherwise amplified in the quenched condensate~\cite{Raman17}. In the Bogoliubov analysis of the initial axial polar state at $q_f<0$ for $B'=0$, the characteristic time and length scales for the quench dynamics are given by $t_q\propto|q_f|^{-1/2}$ and $l_q\propto|q_f|^{-1/2}$, respectively, which was also experimentally demonstrated~\cite{Kang17}. Then, one may consider a dimensional analysis to set the characteristic dephasing condition as $\mu_B B'l_q\sim h/t_q$. This suggests $|q_{th}|\propto B'$, which cannot explain the measured exponent $\alpha$ in the experiment.

We also investigate the dependence of $q_{th}$ on the sample size for various $B'$ [Fig.~5(b)]. For a given $B'$, it is observed that $|q_{th}|$ increases noticeably as $N_{tot}$ decreases below a certain number which is reduced for lower $B'$. This is consistent with the previous observation in Fig.~5(a) that the condensate with the smaller $N_{tot}$ exhibits the larger exponent $\alpha$ and takes off earlier from the power-law line. The measurement results show that the finite size effect enhances the metastability of the axial polar state, but it is not a fundamental requisite for having the metastable state. 

It is important to note that the quench threshold for $B'\approx 0$ was measured to be $q_{th}/h\approx-0.7$~Hz regardless of the sample size [Fig.~5(b)]. This is different from the general expectation that $q_{th}$ should be equal to the critical point $q_c=0$ at $B'=0$. We attribute its deviation to a residual field gradient $B_z'$, along the tightly confining $z$ direction. In the experiment, it is difficult to precisely calibrate $B'_z$ using the interferometric method, due to the small sample extent along the $z$ direction. From the power-law relation of $|q_{th}|$ with $B'$ in Fig.~5(a), the residual field gradient is estimated to be $B'_z < 0.2$~mG/cm.

\subsection{Quantum critical point}

The quantum critical point between the axial and planar polar phases is fundamentally given by $q_c=0$, where the spin rotation symmetry is fully restored. However, in Ref.~\cite{Raman17} it was reported that the critical point is located at $q_c/h\approx+0.65~$Hz. The physical origin of its deviation from the ideal point was not identified. Furthermore, the shift direction is positive in $q$, which cannot be accounted for by the field gradient effect observed in this work.

To address the issue of locating the critical point, we examine the ground-state spin profile of the condensate as a function of $q_f/h$. We prepare the ground state by rotating the spin direction $\boldsymbol{d}$ to the $xy$ plane by applying a pulse of RF field before quenching to a target $q_f$. The initial condensate is an equal mixture of the $m_z=\pm 1$ spin components and when $q_f<q_c$, the system can easily relax to the ground state via spin flow dynamics~\cite{Kim17}. If $q_f>q_c$, it will be revealed by the $m_z=0$ population in the ground-state condensate. In Fig.~6, we display $\eta$ measured at $t=3$~s after the quench as a function of $q_f/h$. The hold time $t=3$~s is set to be long enough to ensure the equilibrium state. For $B'\approx0$, a sudden onset of $\eta$ is observed when $q_f/h$ increases over $\approx -0.7$~Hz, whose position is coincident with the measured threshold $q_{th}$ for the collapse of the axial polar state. This is surprising because it means that the axial polar state is the true ground state of the system for $q/h>-0.7$~Hz, i.e., $q_c/h\approx-0.7$~Hz.

\begin{figure}[t]
	\includegraphics[width=8.5cm]{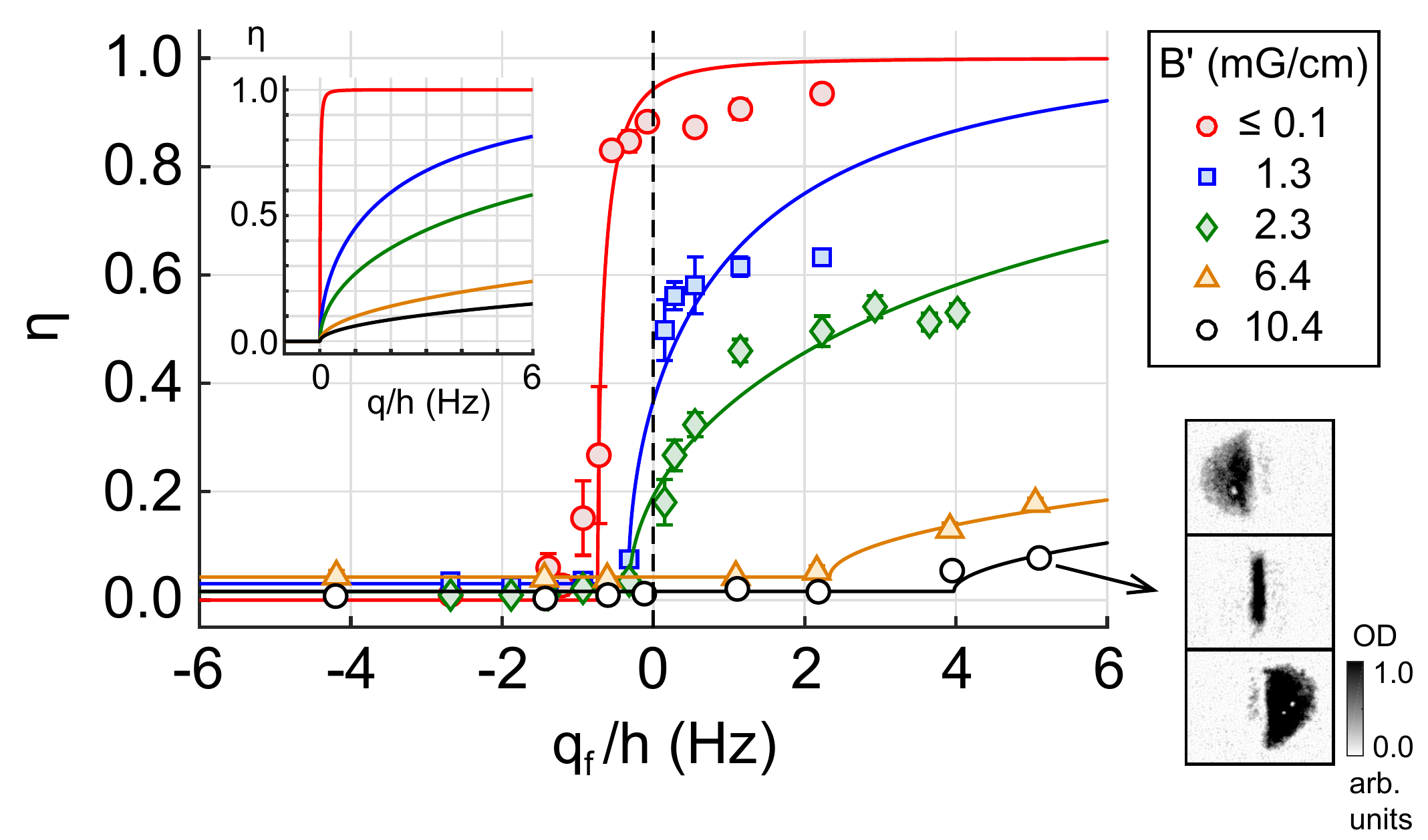}
	\caption{Fractional population $\eta$ of the $m_z=0$ component measured at $t=3$~s after the quench as a function of $q_f/h$ for various $B'$ ($N_{tot}\approx 4.5\times 10^6$).	The condensate was initially prepared in the easy-plane polar state. The solid lines show curve fits of $\eta(q_f)=\eta_{m}(q_f -q_{th}^*, B'_\textrm{eff})$ to the measurement data sets, where $\eta_{m}(q,B')$ is the analytic fucntion of $\eta$ for the ground state in the mean-field description of Sec.~II. As free parameters, $q_{th}^*$ represents the onset point of $\eta\neq 0$ and $B'_\textrm{eff}$ is the effective magnetic field gradient for the response of $\eta$. Each data point was typically obtained from five measurements and its error bar indicates the standard error of the mean. The left inset shows $\eta_m(q,B')$ for the experimental conditions. The right inset displays absorption images of the $m_z$$=1$ (up), $m_z$$=0$ (middle), $m_z$$=-1$ (down) spin components of the condensate after 3~s relaxation at $q_f/h\approx 5.1$~Hz under $B'\approx 10.4$~mG/cm. The density-depleted holes in the $m_z=\pm1$ components indicate quantum vortices generated via complex spin flow dynamics in the relaxation from the inital easy-plane polar state~\cite{Kim17}.}
\end{figure}

This finding seems to be inconsistent with our previous explanation of $q_{th}\neq 0$ for $B'\approx 0$, where we attributed it to the residual field gradient $B'_z$ along the $z$ direction. To test the possible association of $q_c$ not being zero with $B'_z\neq 0$, we repeat the same measurements with various magnetic field configurations, where the background magnetic field $B_z$ is slightly changed or reversed while we maintains the field gradient $B'<0.1~$mG/cm in the $xy$ plane; but, $B'_z$ remains uncontrolled. In our measurements, we observe that $q_{th}$ varies even by a few Hz in the $q<0$ side and, also, that the onset point $q_{th}^*$ for $\eta\neq 0$ in the ground state always follows the variations of $q_{th}$. Thus, it is tentatively suggested that the field gradient along the tightly confining direction affects the quantum critical point of the system, although there is currently no theoretical scenario supporting the possibility.

Another notable observation in the $\eta$ measurement in Fig.~6 is that as $B'$ increases, $q_{th}^*$ is shifted from $q_c$ to the positive $q$ direction. This might be attributed to the domain wall energy, which is neglected in the mean-field description of the ground state. Under a field gradient, an axial polar domain of the $m_z=0$ component is sandwiched by ferromagnetic domains and its spatial extent decreases with increasing $B'$ (Fig.~6 inset). Therefore, for a given energy cost of the domain walls, it can be energetically unfavorable with small $q>0$ to have the middle axial polar domain. A quantitative analysis on the energetics of the spin domain structure will be pursued in future works. We note that the shift of $q_{th}^*$ can be ensured by observing spin domain formation via thermal cooling at a fixed $q$, which might provide a more reliable method to reach the ground state of the system~\cite{Gerbier12,Liu14,Gerbier17,Gerbier18}.

\section{Summary and Outlook}

\begin{figure}[t]
	\includegraphics[width=8.5cm]{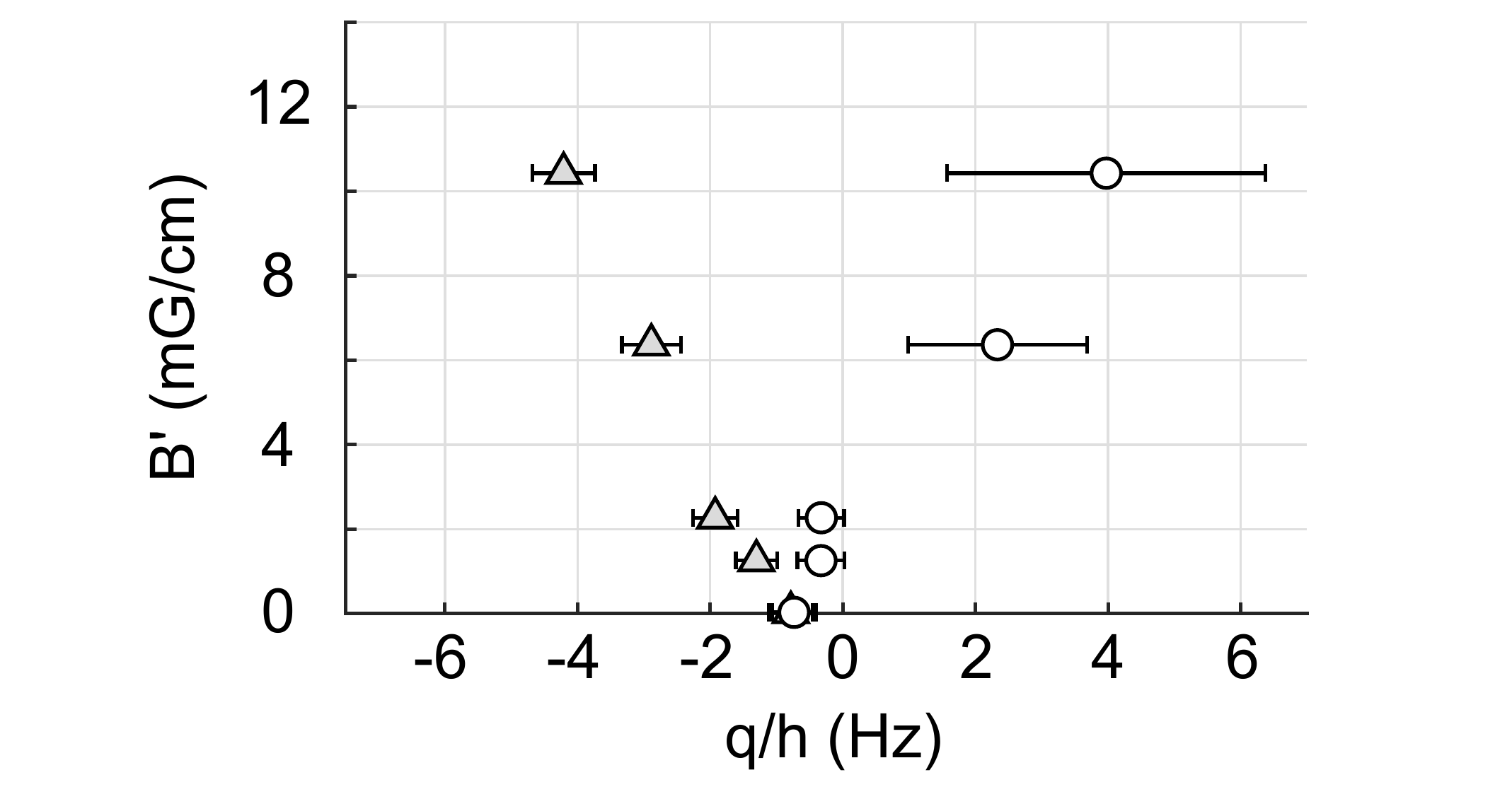}
	\caption{Measurement results of $q_{th}$ and $q_{th}^*$ in the plane of $q/h$ and $B'$. $q_{th}$ is the threshold for the metastable axial polar state to collapse, and $q_{th}^*$ is the threshold for the ground state to have the $m_z=0$ spin compomponent. The data points are obtained from the fits to the $\eta$ measurement data in Fig.~4 and Fig.~6 (see the figure captions).}
\end{figure}

We observed the metastable hard-axis polar state in a spin-1 antiferromagnetic spinor BEC under a magnetic field gradient. The metastability was shown to be driven by the field gradient and break down below a certain threshold $q_{th}$ of the quadratic Zeeman energy, due to the intrinsic dynamical instability. We investigated the dependence of the threshold $q_{th}$ on the field gradient $B'$, and found that $q_{th}$ is highly sensitive to $B'$ in the vicinity of $B'\approx 0$, exhibiting power-law behavior of $|q_{th}|\propto B'^{\alpha}$ with $\alpha \sim 0.5$. The finite size effect on $q_{th}$ was also examined, showing that the metastability is enhanced for small samples. By exploring the ground spin states of the condensate depending on $q$, an evidence was found that the quantum critical point is shifted away from $q_c=0$. In Fig.~7, we summarize the measurement results of $q_{th}$ for small $B'$, together with $q_{th}^*$ for the onset of $\eta\neq 0$ in the ground state.

We expect that our measurement results will stimulate further theoretical investigations on the underlying mechanism of the observed field gradient effect in the quantum phase transition dynamics in the spinor condensate. In particular, the observation of the shift of the quantum critical point poses an open question as to how the field gradient along the tight confining direction of the sample can affect the critical point of the system. This question is crucial for investigating possible scaling behavior in the quantum transition dynamics~\cite{Kang17,Blakie17:2} and studying many novel phenomena anticipated near the quantum critical point, such as the formation of spin-singlet states~\cite{Yip00,Baym06,Julia-Diaz16,Liu18} and critical spin superfluidity~\cite{Kim17,Ferrari18,Wu15,Sonin18}.

\begin{acknowledgments}
We thank Uwe Fischer and Seongho Shin for discussions. This work was supported by the Samsung Science and Technology Foundation under Project Number SSTF-BA1601-06.
\end{acknowledgments}

\end{document}